\documentclass[a4paper]{article}

\newcommand{\convMatrix}[3]{
\begin{math}
\begin{bmatrix}
\text{conv},\ 1 \times 1,\ #1 \\
\text{conv},\ 3 \times 3,\ #1 \\ 
\text{conv},\ 1 \times 1,\ #2 
\end{bmatrix} \times\ #3
\end{math}
}

\newcommand{\convRow}[4]{$ \text{#1},\ #2 \times #3,\ #4 $}

\usepackage{INTERSPEECH2019}
\usepackage{multirow}

\usepackage{xcolor} 

\title{A Deep Neural Network for Short-Segment Speaker Recognition}
\name{Amirhossein Hajavi$^1$, Ali Etemad$^{1,2}$}
\address{$^{1}$Department of Electrical and Computer Engineering, $^2$Ingenuity Labs\\Queens University, Kingston, Canada}
\email{\{a.hajavi, ali.etemad\}@queensu.ca}

\begin{document}
\maketitle

\begin{abstract}
Today's interactive devices such as smart-phone assistants and smart speakers often deal with short-duration speech segments. As a result, speaker recognition systems integrated into such devices will be much better suited with models capable of performing the recognition task with short-duration utterances. In this paper, a new deep neural network, UtterIdNet, capable of performing speaker recognition with short speech segments is proposed. Our proposed model utilizes a novel architecture that makes it suitable for short-segment speaker recognition through an efficiently increased use of information in short speech segments. UtterIdNet has been trained and tested on the VoxCeleb datasets, the latest benchmarks in speaker recognition. Evaluations for different segment durations show consistent and stable performance for short segments, with significant improvement over the previous models for segments of 2 seconds, 1 second, and especially sub-second durations (250 \textit{ms} and 500 \textit{ms}).
\end{abstract}


\section{Introduction}
Speaker recognition has seen profound improvements due to the recent advancements in deep learning. Accordingly, accuracy levels of proposed deep neural networks (DNN) for speaker recognition (both verification and identification) are far surpassing previous state-of-the-art techniques. Recent examples include the use of embeddings obtained from convolutional neural networks (CNN) for speaker recognition in \cite{nagrani_VoxCeleb, cai2018analysis, chung_voxceleb}, the use of auto-encoder models for speaker identification in \cite{zhang_deep_2015, tirumala_deep_2017}, and a number of cases utilizing ResNet for both speaker recognition and identification in \cite{novoselov2018triplet, yadav2018learning}.

As discussed in the review \cite{reynolds_overview_2002}, initial attempts at speaker recognition were performed under highly controlled conditions with very limited size of vocabulary. More recently, challenging factors such as environment noise, impersonation, and different ethnic diversities, are being tackled using techniques such as UBM-GMM and Joint Factor Analysis \cite{kenny_joint_2006}, as well as hybrid GMM and support vector machines \cite{togneri_overview_2011}. Emergence of challenging datasets such as Speaker In The Wild (SITW) \cite{mclaren_speakers_2016} and its extended variations such as VoxCeleb1 and VoxCeleb2 \cite{nagrani_VoxCeleb, chung_voxceleb}, with more than 5,000 speakers and one million utterances, have enabled the opportunity to tackle speaker recognition in real-world scenarios. As a result, a number of deep learning solutions have been proposed for this purpose, including the models studied in \cite{hinton_deep_2012, song_i-vector_2013, yamada_improvement_2013, lei_novel_2014, lei_application_2014, kenny_deep_2014, matejka_neural_2014, lopez-moreno_automatic_2014, ghahabi_deep_2017,shon2018frame, xie2019utterance, park_training_2018, jung_avoiding_2018, wang_centroid-based_2019, kumar_convolutional_2018, nidadavolu_cycle-gans_2019, rohdin_speaker_2019}. 

Current solutions for speaker recognition either rely on the use of an entire utterance for carrying out the recognition task \cite{ghahabi_deep_2017, okabe2018attentive} or require a segment of speech, often two seconds or greater, for accurate performance  \cite{nagrani_VoxCeleb,cai2018analysis,chung_voxceleb, xie2019utterance}. However, with the emergence of voice-based interactive devices such as smart-phone assistants, smart home devices such as smart speakers, in-vehicle entertainment and navigation systems, and other consumer electronics, it is imperative that accurate speaker recognition be performed using short-duration speech segments. Despite the fact that most existing speaker recognition solutions target medium or long utterances, namely 6 seconds or more, the mentioned target environments often deal with short-duration commands, such as "Hey Siri", "Okay Google", "Volume Up", or others. As a result, the aim of this work is to develop a DNN architecture capable of accurate speaker verification with short-segment utterances, specifically under 2 seconds.

In this study, a novel architecture has been proposed to tackle the problem of speaker verification using short-duration segments. Our proposed model is applied over a windowed duration of utterance, where very short segments can be used as inputs. The model creates an embedding of the received utterance, which can be accurately verified with respect to different individuals. Our proposed architecture is designed with the aim of preserving the information often lost through DNN pipelines by means of utilizing the information at multiple junctions and feeding that information to three fully connected networks (FCN). We use the VoxCeleb2 dataset for training, and evaluate our model with respect to state-of-the-art techniques such as \cite{chung_voxceleb}, \cite{xie2019utterance}. 

The rest of the paper is structured as follows. First, we go through the related work, followed by a detailed description of our proposed model architecture. Next, we evaluate the performance of the proposed method with respect to recent works in the field. 


\section{Related Work}
DNNs were first used for speaker recognition in \cite{hinton_deep_2012}, and have since shown promising results by outperforming the traditional HMM-GMM techniques. Since the introduction of DNNs, attempts have made to incorporate such techniques with I-Vector methods in \cite{song_i-vector_2013, kenny_deep_2014}. Later on, further attempts were made to tackle speaker recognition with DNN under difficult conditions such as distant talking \cite{yamada_improvement_2013, lei_novel_2014}. This was followed by the integration of I-Vector with DNNs \cite{kinnun_et_al2017}, which outperformed HMM-GMM-based methods, which were considered state-of-the-art at the time.

Next, novel deep neural network architectures such as auto-encoders \cite{zhang_deep_2015, tirumala_deep_2017} found their way into speaker recognition. In these methods, auto-encoders were used to create embeddings in a lower dimension prior to reconstruction. These lower dimension embeddings, in the form of bottle-neck features, were used to discriminate speakers.

\begin{figure*}[ht]
  \centering
  \includegraphics[width=1.0\linewidth]{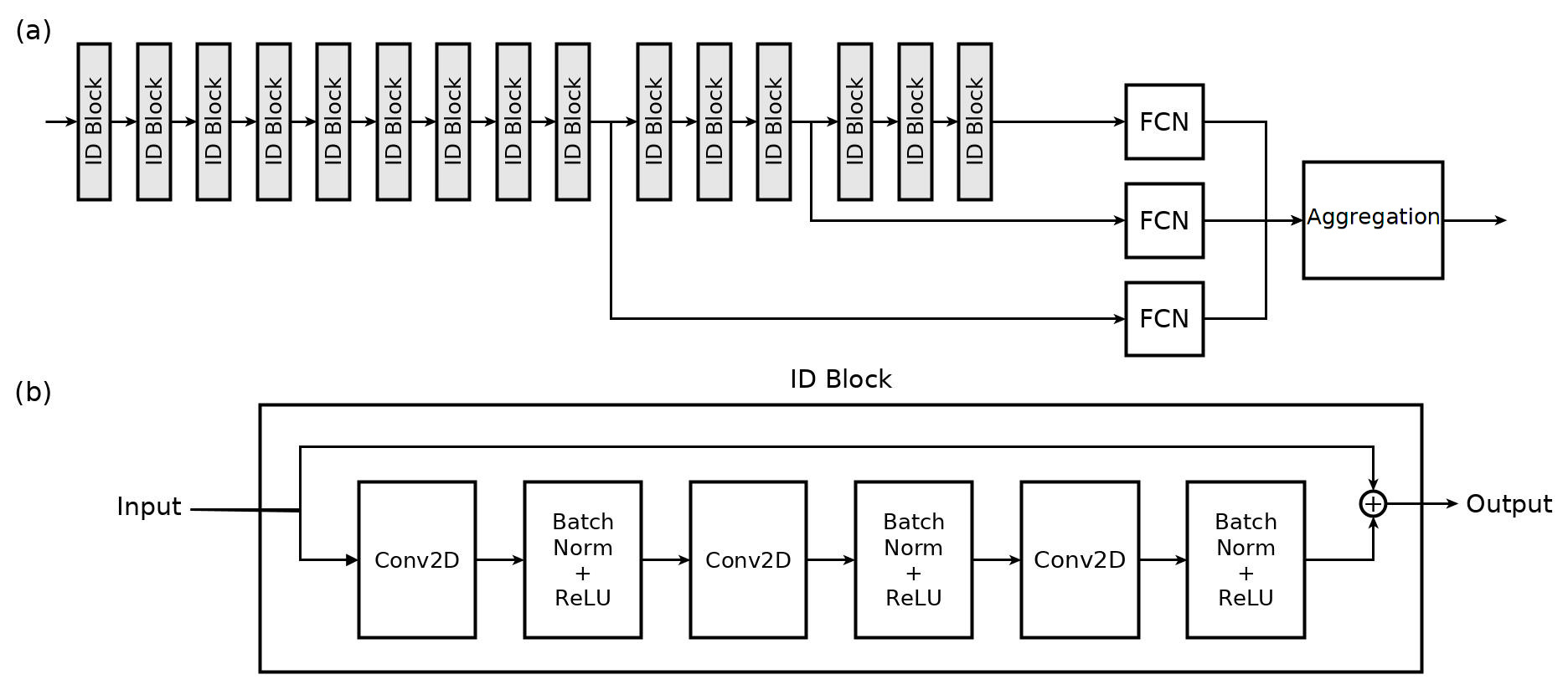}
  \caption{Architecture of our proposed model, UtterIdNet: (a) the overall scheme, followed by (b) the internal architecture of ID Blocks.}
  \label{fig:scheme}
\end{figure*}

Recurrent neural networks (RNN) have been utilized in a number of studies. Recently, RNN models were employed in \cite{park_training_2018, jati_unsupervised_2018, wang_centroid-based_2019} with mel-frequency cepstrum coefficients (MFCC) as inputs. In \cite{wang_centroid-based_2019}, a long short-term memory (LSTM) architecture was applied on MFCC, resulting in an embedding used to verify the speaker of the utterance by means of cosine distance. Other attempts, such as the model proposed in \cite{park_training_2018} and \cite{jati_unsupervised_2018}, have used the LSTM architecture as an intermediate tool in extracting i-vectors. More sophisticated models combining CNN and RNN-based solutions were proposed in \cite{kumar_convolutional_2018} and \cite{ jung_avoiding_2018}, applying several convolution layers in between the MFCC input and the RNN. 

Utilizing adversarial networks has also been explored for speaker recognition. Proposed solutions in \cite{ding_mtgan, fang_channel_2019, nidadavolu_cycle-gans_2019, nidadavolu_cycle-gans_2019, rohdin_speaker_2019} explore the use of adversarial networks and generative adversarial networks both as discriminative models for verification of the speaker as well as generative models. Such generative models were mostly used to transform the conditions of the utterance into more convenient environments in which to perform the speaker recognition task. 

Many studies have utilized CNN-based models for speaker recognition. In \cite{nagrani_VoxCeleb} a CNN architecture was introduced, which outperformed I-Vector-based methods with a large margin. Using convolution and AveragePool/MaxPool layers, the model was able to achieve an improvement in accuracy of approximately 10\%. More advanced CNN-based architectures were then used for speaker recognition. For example, considering successful performance in face recognition tasks, VGGnet \cite{yadav2018learning} was used to tackle both identification and verification, outperforming simple CNN architectures. The ResNet architecture \cite{cai2018analysis, chung_voxceleb, xie2019utterance} was also utilized for speaker recognition, achieving the state-of-the-art by outperforming the previous VGGnet models. Continuing the trend of employing successful models from face recognition, a ResNet architecture proposed in \cite{xie2019utterance} utilizing novel aggregation methods such as GhostVlad and NetVlad, outperformed the previous use of a 50 layer ResNet model \cite{chung_voxceleb}. 

\section{Method}
In this section, the feature extraction process and our proposed model are described:

\subsection{Feature extraction} 
To extract informative frequency features from the short-segment utterances, Short-Time Fourier Transform (STFT) is used in this paper. STFT, as shown in Equation~\ref{eq:stft} is the Fourier transform of the signal under a time window of $\tau$. In Equation~\ref{eq:stft}, the $w(t-\tau)$ is the window operation applied over the signal to capture the frequency features within that specific duration. We used a window of 25 \textit{ms}, with a stride of 10 \textit{ms}, similar to \cite{nagrani_VoxCeleb, cai2018analysis, chung_voxceleb, xie2019utterance}.
\begin{equation}
  X(t,f) = \int^\infty_{-\infty}w(t-\tau) x(\tau) e^{-j 2 \pi f \tau} d\tau
  \label{eq:stft}
\end{equation}

\subsection{Model Architecture} 
Our proposed model is inspired by two highly popular and successful deep neural network architectures, namely ResNet \cite{he_deep_2016} and DeepID3 \cite{sun_deepid32015}. The main contribution made in \cite{sun_deepid32015} was driven by the idea of preserving information often filtered out through MaxPool layers in traditional CNN models. The solution provided in the DeepID network, and later in the enhanced DeepID3, was to feed the outputs of some of the MaxPool layers to an FCN, accompanied by a softmax layer. In their model, a voting mechanism was then used to aggregate the outputs of the softmax layers. The ResNet architecture, on the other hand, reformulated the layers as learning residual mechanism given the layer inputs. These residual networks were shown to be trained easier, while performing more accurately given the increase in depth.

Speaker recognition with short-segment utterances requires maximizing the use of the limited amount of information within the available input window. Furthermore, the large scale of the VoxCeleb2 dataset requires learning methods to be efficient in terms of training epochs and resources. As a result, we decided to use the properties of the aforementioned models to create a new architecture for the task at hand. 

Our proposed architecture, UtterIdNet, is presented in Figure~\ref{fig:scheme} (a). Similar to the DeepID3 model, the data flow has been branched in two other location besides the output of the final block. The internal architecture of an Identity (ID) Block can be seen in Figure~\ref{fig:scheme} (b). The ID Blocks in our model transform the input spectrograms to a level of individual-specific embeddings, followed by the FCNs for a final embedding transformation. Lastly, an aggregator applies a non-linear voting scheme over the FCN outputs. Details such as number of filters and kernel sizes are summarized in Table~\ref{tab:dsd_scheme}. 

The input utterance segments go through 9 ID Blocks to reach the required level of speaker-specific discrimination. This depth was determined though empirical evaluation. The information achieved at this point is preserved through training an FCN through a branched route. Next, successive to three new ID Blocks, another similar FCN is trained, followed by the final three ID Blocks which lead to the last FCN. All the FCNs are trained using an Adam optimizer \cite{kingma2014adam}. Once the utterances have gone through the entire UtterIdNet pipeline, as discussed, three embedding vectors are generated by the FCNs, which contain discriminative speaker-specific features. 

\subsection{Aggregation}
A non-linear aggregator combines the outputs of the three FCNs to produce a 512 dimensional representation of the individuals within the dataset. The non-linear combination used to aggregate the FCN outputs is provided in Equation~\ref{eq:tanh}, 
\begin{equation}
  \rho = tanh(W_1 \times Em_1 + W_2 \times Em_2 + W_3 \times Em_3),
  \label{eq:tanh}
\end{equation}
where $W_1$, $W_2$, and $W_3$ represent weight vectors, and $Em$ denotes the FCN embeddings. In this paper these weights are set to be trained for each dimension within the embeddings individually through an Adam optimizer with a softmax function. If all the indices within each $W$ were selected using a uniform function, our aggregator would act as the same voting mechanism originally implemented for DeepId3 \cite{sun_deepid32015}.

Successive to extraction of embeddings for each short utterance segment, a simple aggregation scheme, Time-Distributed Voting (TDV) is utilized. This aggregation technique collects embedding vectors for successive durations of the input utterance, and carries out a voting mechanism to verify the speaker identity. The reason for utilizing a simple aggregator was the fact that our solution is aimed at speaker recognition for short-segment utterances. Hence, it is likely that by employing more advanced aggregation techniques, better performance is achieved on longer utterance segments.

\begin{table}[t]
\caption{Architectural details of our proposed model.}
\label{tab:dsd_scheme}
\resizebox{\columnwidth}{!}
{
\begin{tabular}{c|c|c}
\hline
Module & Input Spectrogram $(257 \times T \times 1)$ & Output Size \\[.1cm] \hline
\multirow{16}{*}{ConvModule}
& \convRow{Conv2D}{7}{7}{64} & $ 257 \times T \times 64 $ \\[.1cm]
\cline{2-3} & \convRow{Maxpool}{2}{2}{\text{stride}\ (2,2)} & $ 128 \times T/2 \times  64 $ \\[.1cm]
\cline{2-3} & \convMatrix{48}{96}{2} & $ 128 \times T/2  \times  96 $ \\[.1cm]
\cline{2-3} & \convMatrix{96}{128}{3} & $  64 \times T/4  \times 128 $ \\[.1cm]
\cline{2-3} & \convMatrix{128}{256}{3} & $  32 \times T/8  \times 256 $ \\[.1cm]
\cline{2-3} & \convMatrix{256}{512}{3} & $  16 \times T/16 \times 512 $ \\[.1cm]
\cline{2-3} & \convRow{Maxpool}{3}{1}{\text{stride}\ (2,2)} & $ 7 \times T/32 \times 512 $ \\[.1cm]
\cline{2-3} \hline Embedding & FCN $\times$ 3, 512 & 1526 \\[.1cm]
\cline{2-3}
\hline Nonlinear  & \multirow{2}{*}{FCN, 512} & \multirow{2}{*}{512}\\[.1cm]
aggregation &  & \\ 
\cline{2-3}
\hline
\end{tabular}
}
\end{table}

\begin{table*}[t]
  \caption{The performance of our model, UtterIdNet, with respect to \cite{xie2019utterance}, trained with VoxCeleb2 and tested on the VoxCeleb1 dataset. Our model performs consistently as the segment durations are decreased, outperforming \cite{xie2019utterance} with considerable margins for sub-second segments.}
  \label{tab:results_sota}
  \centering
  \begin{tabular}{l l l l l l l}
    \toprule
      & \multirow{2}{*}{\textbf{Model}} & \multirow{2}{*}{\textbf{Aggregation}} &  \textbf{EER\% } & \textbf{EER\%} & \textbf{EER\% } & \textbf{EER\%}  \\
      & & \textbf{ }  & \textbf{250 \textit{ms}}  & \textbf{500 \textit{ms}}  & \textbf{1 \textit{sec}} & \textbf{2 \textit{sec}} \\

    \hline\hline
    & & & & & & \\[-.2cm]
    Xie et al. \cite{xie2019utterance} & Thin ResNet34 & GhostVlad & 23.23 &  10.58 & 9.25 & 7.97 \\
    & & & & & & \\[-.2cm] 
    \textbf{Ours} & \textbf{UtterIdNet} & \textbf{TDV} & \textbf{6.88} &  \textbf{6.46} & \textbf{6.41} & \textbf{6.33}\\
    \bottomrule
  \end{tabular}
\end{table*}

\begin{table*}[t]
  \caption{The result of UtterIdNet for full-length utterances. For longer segments, our model is outperformed by \cite{okabe2018attentive}, \cite{chung_voxceleb}, and \cite{xie2019utterance}.}
  \label{tab:results_full}
  \centering
  \begin{tabular}{l l l l l l}
    \toprule
      & \textbf{Model} & \textbf{Loss} &  \textbf{Dims } & \textbf{Train set } & \textbf{EER\% } \\

    \hline\hline
    & & & & \\[-.3cm] 
    
    Nagrani et al. \cite{cai2018analysis} & I-Vector + PLDA & -- & -- & VoxCeleb1 &  8.80 \\[.1cm]
    
    Cai et al. \cite{cai2018analysis} & ResNet34 + SAP & A-softmax + PLDA & 128 & VoxCeleb1 & 4.40 \\[.1cm]
    
    Cai et al. \cite{cai2018analysis} & ResNet34 + LDE & A-softmax + PLDA  & 128 & VoxCeleb1 & 4.48 \\[.1cm]
    
    Okabe et al. \cite{okabe2018attentive} & TDNN (X-Vector) + TAP & softmax & 1500 & VoxCeleb1 & 3.85 \\[.1cm]
    
    Hajibabai et al. \cite{hajibabaei2018unified} & ResNet20 + TAP& AM-softmax & 128 & VoxCeleb1 &4.30 \\[.1cm]
    
    Chung et al. \cite{chung_voxceleb} & ResNet50 + TAP & softmax + Contrastive & 512 & VoxCeleb2 & 4.19 \\[.1cm]
    
    Xie et al. \cite{xie2019utterance} & Thin ResNet34 + TAP & softmax & 512 & VoxCeleb2 & 10.48 \\[.1cm]
    
    Xie et al. \cite{xie2019utterance} & Thin ResNet34 + GhostVlad & softmax & 512 & VoxCeleb2 & 3.22 \\[.1cm]  
    Ours & UtterIdNet + TDV & softmax & 512 & VoxCeleb2 & 4.26\\[.1cm]
    \bottomrule
  \end{tabular}
\end{table*}

\section{Experiments and Results}
In this section, we describe the datasets and the experimental setup utilized for speaker verification. Next, the results obtained by our model using different short segment durations are presented, and compared to the state-of-the-art \cite{xie2019utterance}. Furthermore, in addition to short-segment windows, the performance of the proposed model on full utterances has been evaluated with respect to recent successful techniques. Lastly, the potential of UtterIdNet for real-time applications is explored, followed by a discussion on required memory resources for potential edge-device applications.

\subsection{Datasets}
The VoxCeleb1 \cite{nagrani_VoxCeleb} and VoxCeleb2 \cite{chung_voxceleb} datasets have been collected through automatic pipelines from open source media, and contain 1,250 and 5,994 speakers respectively. The utilized pipeline was based on computer vision techniques such as face recognition and active speaker detection. Due to the datasets being collected from real media content, \textit{in the wild} scenarios were maintained in these datasets.

VoxCeleb2, as discussed by the authors in \cite{chung_voxceleb}, contains several flaws in its annotations. Therefore, its use in testing of models has not been advised, while it has been widely used for training purposes. VoxCeleb1, however, has been collected under extremely strict constraints, leading the dataset to be free of any annotation issues. As a result, models are often tested on this dataset as a benchmarking practice. Accordingly, in this paper, the VoxCeleb2 with over 1 Million utterances has been used for training, while VoxCeleb1 was used as the test set for our proposed model.

\subsection{Training}
Training of the proposed model was performed using a standard softmax loss. For optimization, an Adam optimizer with a decaying learning rate and an initial value of $10^{-4}$ was utilized. The learning rate decay was a factor of 10 for every 36 epochs. Training was done on a single Nvidia 1080 Ti GPU.

\subsection{Testing and verification}
Table \ref{tab:results_sota} presents the results. In this table, the performance of UtterIdNet with respect to Thin ResNet34 + GhostVlad \cite{xie2019utterance} is provided. In order to evaluate the impact of segment durations on performance, 250 \textit{ms}, 500 \textit{ms}, 1 \textit{sec}, and 2 \textit{sec} segment sizes were used. UtterIdNet outperforms Thin ResNet34 + GhostVlad in all the different short-segment scenarios. Notably, the performance of UtterIdNet shows reasonable amount of consistency and stability as the segment size is decreased. As a result, the proposed UtterIdNet model outperforms ResNet34 + GhostVlad by a considerable margin for sub-second segments and especially for 250 \textit{ms} segments. The performance of other methods such as x-vector \cite{okabe2018attentive} or ResNet50 \cite{chung_voxceleb} have not been reported for short segments. Moreover, to the best of our knowledge, these models were not publicly available, and therefore a direct comparison was not possible.

To further evaluate our method, we also compared the performance on full-utterance inputs. The results of this experiment are presented in Table \ref{tab:results_full}. As illustrated in this table, UtterIdNet is closely outperformed by \cite{okabe2018attentive}, \cite{chung_voxceleb}, and \cite{xie2019utterance}. Nonetheless, given that TDV was kept simple for high efficiency for short-segment utterances, this margin may be reduced or eliminated with more advanced aggregation techniques. 

Using the gradient sharing approach enabled through our proposed method, convergence was achieved at a much quicker rate than that of other recent DNN solutions. This was observed and evident by the performance with respect to the number of training epochs. In \cite{xie2019utterance} it was reported the converged performance was obtained at 48 epochs, while UtterIdNet produced the results in Tables \ref{tab:results_sota} and \ref{tab:results_full} after only 25 epochs. Moreover, for the same number of epochs (25), the full-length utterance performance of \cite{xie2019utterance} appears to be 16\% EER, which is considerably outperformed by UttterIdNet.

\subsection{Identification}
The latest reported identification rate for VoxCeleb1 and VoxCeleb2 datasets has been 89.5\% \cite{yadav2018learning}. This performance was achieved using \textit{full-length utterances}, and on VoxCeleb1 which contained 1,250 speakers. We tested UtterIdNet for identification with 250 \textit{ms} segments on the VoxCeleb2 validation set, which contained 5,994 speakers. Despite shorter segment sizes and a significantly larger number of speakers (approximately 4 times), we achieved an accuracy of 84.3\%. These results show significant promise and the potential for considerable improvement in speaker identification in addition to verification, which we intend to study in future works.

\subsection{Memory requirements}
In order for the proposed model to be utilized in edge nodes such as smart-phones and Internet of Things (IoT) devices, certain memory constraints will need to be met. Given the fact that our model consists of 3 FCN components and 15 ID Blocks, to address the concerns regarding the memory requirements of such a system and the feasibility of its implementation in the described context, we evaluated the required memory resources during run-time. The obtained estimated memory need for UtterIdNet was 268 \textit{MB}. This amount of required memory is feasible for many current consumer edge devices, especially in the context of smart-phones, smart homes, and vehicles. As a result, we believe UtterIdNet can be widely employed in these contexts for low-latency recognition and verification of speakers.

\section{Conclusions}
In this paper a new DNN, UtterIdNet, was proposed with the aim of an efficient increase of information use for short speech segments. We evaluated our method using the VoxCeleb datasets, and demonstrated that UtterIdNet outperforms the state-of-the-art for short segments. Specifically, we utilized 250 \textit{ms}, 500 \textit{ms}, 1 \textit{sec}, and 2 \textit{sec} segments, where UtterIdNet showed significant improvement in the sub-second segment range. The efficient learning of information in our proposed model was not only evident by the considerable improvement in accuracy, but also by the fact that in order to train properly, UtterIdNet took approximately half the number of epochs as the state-of-the-art. While UtterIdNet was outperformed by a small margin in full-utterance segments, we believe the choice of the simple aggregation technique for combining the different short segments within the full utterance could be a contributing factor, which we intend to investigate in future work.

\section{Acknowledgements} The authors would like to thank IMRSV Data Labs for their support of this work. The authors would also like to acknowledge the Natural Sciences and Engineering Research Council of Canada (NSERC) for supporting this research (grant number: CRDPJ 533919-18).

\bibliographystyle{IEEEtran}

\bibliography{mybib}

\end{document}